\documentclass[aps,prd]{revtex4}

\begin{document}
\title{ Neutrino mixing matrix in the 3-3-1 model with heavy leptons and $A_4$ symmetry}
\author{Furong Yin}
\email{yfr@itp.ac.cn}
\address{Institute of Theoretical Physics, Chinese Academy of
Science.\\
P.O.Box 2735, Beijing 100080, China.\\
Graduate School of the Chinese Academy of Science.}

\begin{abstract}
We study the lepton sector in the model based on the local gauge
group $SU(3)_c\otimes SU(3)_L\otimes U(1)_X$ which do not contain
particles with exotic electric charges. The seesaw mechanism and
discrete $A_4$ symmetry are introduced into the model to understand
why neutrinos are especially light and the observed pattern of
neutrino mixing. The model provides a method for obtaining the
tri-bimaximal mixing matrix in the leading order. A non-zero
mixing angle $V_{e3}$ presents in the modified mixing matrix.
\end{abstract}

\maketitle

\section{Introduction}\label{intro}

There is convincing evidence for solar and atmospheric neutrino
oscillations~\cite{nos}. And the experimental results of
Super-Kamiokande \cite{superK}, KamLAND \cite{kam} and SNO
\cite{sno} confirm that neutrinos have small but non-zero masses and
oscillate. The current experimental data are consistent with so
called the tri-bimaximal form \cite{V1}\cite{V2} which, apart from
phase redefinitions, is given by,
\begin{eqnarray}
V_{tri-bi}=\left(
\begin{array}{ccc}
\frac{2}{\sqrt{6}}       &\frac{1}{\sqrt{3}}  &0\\
-\frac{1}{\sqrt{6}}      &\frac{1}{\sqrt{3}}  &\frac{1}{\sqrt{2}}\\
-\frac{1}{\sqrt{6}}      &\frac{1}{\sqrt{3}}  &-\frac{1}{\sqrt{2}}
\end{array}\right)\label{eq:1}
\end{eqnarray}

The explanation of the smallness of the neutrino masses and the
profile of their mixing as required by recent experiments have been
a great puzzle in particle physics. Since in the successful Standard
Model(SM), only the massless neutrinos which pair charged leptons in
three left-handed flavor generations are considered, it appears
obviously that massive neutrinos can be regarded as the definitely
signature of new physics beyond SM.

The neutrinos may acquire naturally small Majorana masses through
the effective dimension-five operator $O_5$ \cite{o5}. For the
standard $SU(2)_L\otimes U(1)_Y$ gauge model, the realizations of
this operator at tree and one loop level were already investigated
in the Ref \cite{neu}, one of the tree level realizations is of
course the canonical seesaw mechanism \cite{seesaw} with one heavy
right-handed neutrino $N_R$ for each $\nu_i$, whereas the new
particles required in the tree-level realizations are most likely
too heavy to be observed experimentally in the near future. In order
to reduce the scale of new physics to be only a few TeV and thus be
observable at future accelerators, another higgs doublet with a
naturally small VEV  is introduced into the model \cite{neu}.

On the other hand, it is an interesting challenge to formulate
dynamical principles that can lead to the tri-bimaximal mixing
pattern given by Eq. (\ref{eq:1}) in a completely natural way as a
first approximation, and many theoretical efforts have been made to
produce such a mixing pattern \cite{tribim}-\cite{A4hv}. For some
years Ma \cite{A4a} has advocated choosing $A_4$, namely, the
symmetry group of the tetrahedron as a family group. In a number of
interesting papers with various collaborators, Ma has shown that a
broken flavour symmetry based on the non-Abelian discrete group
$A_4$ appears to be particularly fit for this purpose
\cite{A4a}-\cite{A4hv}. This non-Abelian discrete finite group
admits one three-dimensional representation $\underline{3}$ as well
as three one-dimensional representations $\underline{1}$
$\underline{1}'$ and $\underline{1}''$, appears simplest discrete
symmetry perfect for 3 families. In most original $A_4$ models, the
SM left-handed leptons $l_L$ and right-handed charged leptons $l_R$
transform as $\underline{3}$ and ($\underline{1}\oplus
\underline{1}' \oplus \underline{1}''$), respectively, or in
opposition. The SM singlet right-handed neutrinos $\nu_R$ which
transform as $\underline{3}$ or ($\underline{1}\oplus \underline{1}'
\oplus \underline{1}''$) under $A_4$ are introduced into the model
for obtaining see-saw neutrinos masses and at the same time
preserving the $SU(2)_L\otimes U(1)_Y$ gauge symmetry. With these
representations, the $A_4$ models can naturally obtain the
tri-bimaximal mixing at first approximation \cite{A4c}.

Here we would like to extend the above application to models based
on the local gauge group $SU(3)_c\otimes SU(3)_L\otimes U(1)_X$
(hereafter 3-3-1 model) with a corresponding enlargement of fermion
representations \cite{pisano}. Because of the two important features
that the number of family is related by anomaly cancellation to the
number of colors, and the third family is treated differently from
the first and second families the 3-3-1 model has received much
attentions \cite{pisano}-\cite{la}. With experimental establishment
of the neutrino oscillations a number of paper has been published to
discuss the neutrino masses and mixing patterns in the model
\cite{foot}. In this paper, we introduce the canonical seesaw
mechanism and $A_4$ symmetry into the framework of one specific
3-3-1 model and study the neutrinos masses and mixing matrix. We
show that in this model, the masses of neutrinos and charged leptons
are generated by two separate scalar sectors, and the neutrinos can
naturally obtain the small masses. In the leading order the neutrino
mixing is just the tri-bimaximal matrix, and after a simple
modification, the non-zero mixing parameter $V_{e3}$ is present.

The paper is organized as follows: In Section \ref{331}, we give a
brief review of the 3-3-1 models and define the framework of our
work. In Section \ref{a4}, the $A_4$ symmetry is introduced into the
model, and the mass mechanisms and mixing matrix of leptons are
represented. Section \ref{mod} discusses the modified neutrino
masses and mixing matrix. The conclusions are given in Section
\ref{con}. Appendix states the basic of $A_4$ symmetry and the
potentials which can give the VEV form of the scalars we used in the
paper.

\section{The 3-3-1 model with the singlet right-handed
Neutrinos}\label{331}

There are different versions of 3-3-1 model. They are nicely
reviewed in ref \cite{yasue}. Consider the electric charge
associated with the unbroken gauge symmetry $U(1)_Q$ which is
defined in general as a linear combination of the diagonal
generators of the group,

\begin{eqnarray}
\widehat{Q}=\widehat{T}_3+\frac{2}{\sqrt{3}}b\widehat{T}_8+X\widehat{I}_3,
\end{eqnarray}
 and then
\begin{eqnarray}
\widehat{Y}=\frac{2}{\sqrt{3}}b\widehat{T}_8+X\widehat{I}_3,
\end{eqnarray}
where $T_a$ ($a=1,2,\cdots, 8$) are the eight generators of
$SU(3)_L$ and $I$ is the unit matrix.

The value of the $b$ parameter determines the fermion assignment and
it is customary to use this number to classify the different 3-3-1
models. Taking $b=\pm 3/2$, for example, we obtain the original
Frampton, Pisano and Pleitez models \cite{pisano}. In this version
the charge conjugation of the right-handed charged lepton for each
generation is combined with the usual $SU(2)_L$ doublet left-handed
leptons components to form an $SU(3)$ triplet $(\nu, e, e^c)_L$. In
the sense that no extra leptons are needed the version can be
considered as minimal. There is no right-handed neutrino in this
minimal version but there are quarks with exotic charges -4/3 and
5/3. As it shown in refs. \cite{foot, la}, if we accommodate the
known left-handed quark and lepton isodoublets in the two upper
components of $3$ and $3^*$ (or $3^*$ and $3$), and forbid the
exotic electrical charges in the possible models, then $b=\pm 1/2$
is mandatory. The original version with right-handed neutrino
\cite{wa}-\cite{dng} is in this category. In this version a
left-handed antineutrino is added to each usual $SU(2)_L$ doublet
left-handed lepton to form a triplet $(\nu, e, \nu^c )_L$.

For our purpose we will consider another version with $b\pm 1/2$.
The $SU(3)_c\otimes SU(3)_L\otimes U(1)_X $ anomaly free fermion
contents can be summarized as follows:

In the lepton sector, we have
\begin{eqnarray}
 &&\varphi_{iL}=\left(
              \begin{array}{c}
                \nu_i\\e_i \\
                E_i
            \end{array}\right)_L \sim (1, 3, -\frac{2}{3}), \,\,\,\,
            N_{iR}\sim  (1, 1,0),\,\,\,\,
   e_{iR}\sim  (1, 1,-1),\,\,\,\, E_{iR}\sim  (1, 1,-1),\,\,\,
   \end{eqnarray}
 where $i = 1, 2, 3$ is a  family index and
$E_{iL/R}$ are negatively charged heavy leptons.  The left-handed
leptons and the charge conjugation of its right-handed counterparts
in this version appear in different multiplets in contrast to the
minimal version \cite{yasue,Lately}  or the version with the
right-handed neutrino \cite{wa}  where all lepton degrees of
freedom, {\it i.e.}, $e_{iL}$, $(e_{iR})^c$, or $\nu_{iL}$,
$(\nu_{iR})^c$ belong to the same triplet. Therefore the present
model has an extra global $U(1)$ symmetry and we can assign a lepton
number for every fields. The introduction of right-handed neutral
Weyl states $N_{iR}$ is optional. The version with no neutral Weyl
states has been studied in refs. \cite{singer,kita}. They are
introduced here is for the tree level realization of the canonical
see-saw mechanism and obviously it does not change the anomaly
cancellation.

In the quark sector, we have
\begin{eqnarray}\nonumber
&& Q_{aL}=\left(
        \begin{array}{c}
           d_a \\
           u_a\\
           U_a
        \end{array}\right)_L\sim (3, 3^*, \frac{1}{3}),\,\,\,\,\
        U_{aR}\sim(3, 1, \frac{2}{3}) \,\,\,\ a=1,2,\,\,\,\,\\\nonumber
&& Q_{3L}=\left(
           \begin{array}{c}
            u_3 \\
            d_3\\
            D_3
          \end{array}\right)_L \sim(3, 3, 0), \,\,\,\,
          D_{3R}\sim(3, 1, -\frac{1}{3})        \\
&&u_{iR} \sim (3, 1, \frac{2}{3}), \,\,\,\,d_{iR} \sim(3, 1,
-\frac{1}{3})\,\,\,\,\,i=1,2,3.
   \label{femion}
   \end{eqnarray}
Note here the five quarks $u_{i R}$ and $U_{a R}$ have the same
quantum number and so are the four quarks $d_{a R}$ and $D_{3R}$.
One can see that the third generation is treated differently from
the first two generations here as we mentioned in the introduction.

We now consider the most general set of scalars that can Yukawa
couple to the above leptons and quarks through either lepton
bilinears, quark bilinears, or quark-lepton bilinears \cite{bb}.
Then all the possible scalar representations under $SU(3)_c\otimes
SU(3)_L\otimes U(1)_X$ are: $(1,1,0)$, $(1,1,-1)$,
$(1,1,-2),(1,3,-\frac{2}{3})$, $ (1,3,\frac{1}{3})$,
$(1,3,\frac{4}{3})$, $(1,6,-\frac{4}{3})$, $(3,1,-\frac{1}{3})$,
$(3,1,\frac{2}{3})$, $ (3,1,-\frac{4}{3})$, $ (3,3,0)$, $ (3,1,1)$,
$(3,3^*,-\frac{2}{3})$, $ (3,3^*,\frac{1}{3})$, $
(3,3^*,\frac{4}{3})$, $(3,6^*,0)$, $(3,6,-\frac{2}{3})$,
$(3,8,-\frac{1}{3})$, $(3,8,\frac{1}{3})$, $(6,1,\frac{1}{3})$,
$(6,1,\frac{4}{3})$, $(6,1,-\frac{2}{3})$, $(6,3,\frac{2}{3})$,
$(6,3^*,0)$, $(6,6,0)$, $(6,8,\frac{1}{3})$, $(8,3^*,-\frac{1}{3})$,
$(8,3^*,\frac{2}{3})$, and their complex conjugates. As for us, in
this work, we choose the scalars to break the symmetry following the
pattern,
\begin{eqnarray}
SU(3)_c\otimes SU(3)_L\otimes U(1)_X\longrightarrow SU(3)_c\otimes
SU(2)_L\otimes U(1)_Y\longrightarrow SU(3)_c\otimes U(1)_Q
\end{eqnarray}
and give, at the same time, masses to the fermion fields in the
model. Then the minimally required scalars are:
\begin{eqnarray}
\chi=\left(
       \begin{array}{c}
        \chi^+ \\ \chi'^0 \\ \chi^0
         \end{array}\right)\sim (1,3,\frac{1}{3}),\,\,\,\,
\rho=\left(
       \begin{array}{c}
         \rho^+ \\ \rho^0 \\ \rho'^0
         \end{array}\right)\sim (1,3,\frac{1}{3}),\,\,\,\,
\eta=\left(
       \begin{array}{c}
         \eta^0 \\ \eta^- \\ \eta'^-
         \end{array}\right)\sim (1,3,-\frac{2}{3}),
\end{eqnarray}
and their complex conjugates. And there may be multiple scalars of
each type, in particular when $A_4$ symmetry is introduced later.
The necessary VEVs are:
\begin{equation}
   \left\langle \chi \right\rangle =
     \left(
        \begin{array}{c}
         0 \\ 0 \\ V
         \end{array}
    \right), \hspace{5mm}
   \left\langle \rho \right\rangle =
       \left(
          \begin{array}{c}
            0 \\v \\0
           \end{array}
       \right)  \label{triplete1}, \hspace{5mm}
   \left\langle \eta \right\rangle =
       \left(
          \begin{array}{c}
             u\\0 \\0
          \end{array}
       \right).  \label{vev}
\end{equation}
here the VEV $V$ is responsible for the first breakdown while $v$
and $u$ are responsible for the second breakdown. So $\chi$ and
$\rho$ have the same quantum numbers but they get VEVs at different
mass scales. Then the scalar $\chi$ breaks $SU(3)_c\otimes
SU(3)_L\otimes U(1)_X$ to $SU(3)_c\otimes SU(2)_L\otimes U(1)_Y$ and
gives large masses to the new fermions as well as non-SM gauge
bosons. The remaining scalars implement $SU(2)_L\times U(1)_Y$
breaking and give the realistic masses to the known fermions and
bosons. Just like the standard $SU(2)_L\otimes U(1)_Y$ gauge model
of Refs. \cite{neu, malatest}, in which in order to acquire
naturally small neutrinos masses, the charged leptons get the masses
from the doublet as in the SM, while the masses of neutral leptons
come from another doublet with a naturally small VEV, in this model,
the changed leptons and neutral leptons get the masses form $v$ and
$u$, respectively, as showed in section \ref{a4}. To keep
consistency with the effectice theory, the VEVs in the model satisfy
the constraint: $V>v>u$.

Notice that flavor changing neutral current (FCNC) in the 3-3-1
models in general is not suppressed naturally either because of the
different treatment to the third generation from the 1st and 2nd
generations of fermions, or the violation of Glashow-Weinberg
natural flavor conservation low \cite{nc}. This is also true to the
present version. This issue has been studied in a number of papers
\cite{montero}, so we will not discuss it further.

\section{ Discrete symmetry $A_4$ and leptons masses} \label{a4}

As discussed in section \ref{intro} the non-abelian discrete group
$A_4$ provides interesting examples of leading to the tri-bimaximal
mixing matrix. The group consists of 12 elements and has 4
irreducible representations ( see refs. \cite{malast,max3} or  the
appendix). Depending on what representations of $A_4$ we choose for
the various fermion and scalar fields there are different schemes.
Here we follow the discussion on the $A_4$ model of the standard
$SU(2)_L\times U(1)_Y$ theory in refs. \cite{malast} and try the
following $ A_4$ assignment to the leptons of the 3-3-1 model
defined in the last section,
\begin{eqnarray}\nonumber
&& \varphi_{iL}=(\nu_i,e_i,E_i)_L^T \sim \underline{3},
         \hspace{26mm} N_{iR}\sim \underline{3},\\
 && e_{1R}\oplus e_{2R} \oplus e_{3R}\sim  (\underline{1}\oplus \underline{1}'\oplus \underline{1}''),
      \hspace{10mm} E^c_{iL}\sim \underline{3}.
\end{eqnarray}

Notice that the SM right-handed charged fermions are assigned to a
$\underline{1}\oplus \underline{1}'\oplus \underline{1}''$ structure
whereas the right-handed neutrinos and heavy right-handed charged
fermions are each given to the $\underline{3}$ representation.

Then $A_4$ and $SU(3)_C \otimes SU(3)_L\otimes U(1)_X$ invariant
Yukawa interactions require to enlarge the scalar sector
correspondingly. Most general scalars now can be:
\begin{itemize}
\item scalars $\rho_i(i=1,2,3)$ which transform as
$\underline{3}$ representation of $A_4$ and Yukawa couple to the
lepton bilinears $\varphi_{iL}$ and $e_{iR}$;
\item scalars $\chi$, $\chi_{1'}$, $\chi_{1''}$ and $\chi_{i}$ which transform as $\underline{1}$,
$\underline{1}'$, $\underline{1}''$, and $\underline{3}$
representations of $A_4$ respectively are needed to Yukawa couple to
the lepton bilinears $\varphi_{iL}$ and $E_{iR}$;
\item scalars $\eta$, $\eta_{1'}$,
$\eta_{1''}$ and $\eta_i$ which transform as $\underline{1}$,
$\underline{1}'$, $\underline{1}''$ and $\underline{3}$
representations respectively and Yukawa couple to the lepton
bilinears $\varphi_{iL}$ and $N_{iR}$,
\end{itemize}
and finally we need $SU(3)_C \otimes SU(3)_L\otimes U(1)_X$ singlet
scalar $\xi$ to generate tree-level Majorana mass term
$M_N(N_R)^cCN_R$.

In practice we find it is sufficient to consider scalars $\rho_i$,
$\chi$, $\eta$, $\eta_i$ and $\xi$. Then the $A_4$ and $SU(3)_c
\otimes SU(3)_L \otimes U(1)_X$ invariant Yukawa interactions in the
lepton sector read,
\begin{eqnarray}\nonumber
L_Y&=&\lambda_1(\overline{\varphi}_{iL} \rho_j)
      e_{1R}+\lambda_2(\overline{\varphi}_{iL}
      \rho_j)''e_{2R}+\lambda_3(\overline{\varphi}_{iL} \rho_j)'e_{3R}
      +\lambda_4(\overline{\varphi}_{iL} E_{jR}) \chi
      +\lambda_5(\overline{\varphi}_{iL} \rho_j E_{kR} )\\
&& +M_N(N_R)^cCN_R+h_1(\overline{\varphi}_{iL}N_{jR})\eta+
    h_2(\overline{\varphi}_{iL}N_{jR}\eta_k)+h.c.
    + \cdot\cdot\cdot  \label{yukawa}
\end{eqnarray}

Here, $(\underline{3}\underline{3})$ transforms as $\underline{1}$,
$(\underline{3}\underline{3})'$ transforms as $\underline{1}'$,
$(\underline{3}\underline{3})''$ transforms as $\underline{1}''$,
$(\underline{3}\underline{3}\underline{3})$ transforms as
$\underline{1}$ under $A_4$ symmetry. We can see that this
interaction actually has a quite simple structure.

From the above Yukawa interaction, the 6 $\times$ 6 mass matrix of
charged lepton ($e_i$, $E_i$) is fund to have the form,
\begin{eqnarray}
m_{eE}= \left(
\begin{array}{cccccc}
\lambda_1v_1 &\lambda_2v_1         &\lambda_3v_1         &0           &\lambda_5v_3    &\lambda_5v_2\\
\lambda_1v_2 &\lambda_2\omega v_2  &\lambda_3\omega^2v_2 &\lambda_5v_3  &0             &\lambda_5v_1\\
\lambda_1v_3 &\lambda_2\omega^2v_3 &\lambda_3\omega v_3  &\lambda_5v_2  &\lambda_5v_1    &0\\
0            &0                    &0                   &\lambda_4V  &0             &0\\
0            &0                    & 0                  &0           &\lambda_4V    &0\\
0            &0                    &0                   &0           &0             &\lambda_4V\\
\end{array}\right)
\end{eqnarray}
where $V$ and $v_i$ are the VEVs for $\chi$ and $\rho_i$
respectively. $v_i$ are taken to be relatively real, and the
numerical subscripts $1,2,3$ of $v$ denote the $A_4$ components, as
in the appendix.

For simplicity we take,
\begin{equation}
v_1 = v_2 = v_3 \equiv v
\end{equation}
Then the mass matrix can be diagonalized by using the unitary
transformations from the weak interaction  eigenstates to mass
eigenstates,
\begin{eqnarray}
 \left(
\begin{array}{c}
e_{i}\\
E_{i}\\
\end{array}\right)_L^w
=U_L^l
 \left(
\begin{array}{c}
e_{i}\\
E_{i}\\
\end{array}\right)_L^m,\,\,\,\,\,\,\,\,\
\left(
\begin{array}{c}
e_{i}\\
E_{i}\\
\end{array}\right)_R^w
=U_R^l
 \left(
\begin{array}{c}
e_{i}\\
E_{i}\\
\end{array}\right)_R^m
\end{eqnarray}
where the $6\times 6$ unitary matrices $U_{L,R}^l$ can be written as
\cite{gtom},
\begin{eqnarray}
 U_L^l=\left(
\begin{array}{cc}
A_L& B_L\\
F_L &G_L\\
\end{array}\right),\,\,\,\,\,\
U_R^l=\left(
\begin{array}{cc}
A_R& B_R\\
F_R &G_R\\
\end{array}\right)
\end{eqnarray}
with
\begin{eqnarray}\nonumber
&& A_L=U(\omega)\cdot \left(
         \begin{array}{ccc}
         c_1& &\\
          &c_2 &\\
         & & c_3
\end{array}\right),\,\,\,\,\,\
   B_L=U(\omega)\cdot \left(
         \begin{array}{ccc}
         s_1& &\\
          &s_2 &\\
         & & s_3
\end{array}\right) \\\nonumber
&& F_L=U(\omega)\cdot \left(
         \begin{array}{ccc}
         -s_1& &\\
          &-s_2 &\\
         & & -s_3
\end{array}\right),\,\,\,\,\,\
  G_L=U(\omega)\cdot \left(
         \begin{array}{ccc}
         c_1& &\\
          &c_2 &\\
         & & c_3
\end{array}\right), \\\nonumber
&& A_R=I_{3\times3},\,\,\,\, B_R=F_R=0, \,\,\,\, G_R=U(\omega)
\end{eqnarray}
\begin{eqnarray}
U(\omega)= \frac{1}{\sqrt{3}}\left(
\begin{array}{ccc}
1&1&1\\
1 & \omega & \omega^2\\
1 & \omega^2 & \omega\\
\end{array}\right).
\end{eqnarray}

Where $c_i=cos\theta_L^{e_i}$ represents the mixing of $e_{iL}$ with
the heavy left-handed charged leptons $E_{iL}$ with,
\begin{eqnarray}\nonumber
tg\theta_1=\frac{2\lambda_4v}{\lambda_5V},\,\,\,\,\
tg\theta_2=-\frac{\lambda_4v}{\lambda_5V},\,\,\,\,\
tg\theta_3=-\frac{\lambda_4v}{\lambda_5V}.
\end{eqnarray}

The charged lepton masses are given by,
$$m_{e_i}=\sqrt{3}\lambda_ic_iv,\,\,\,\,\,\,\,\,\,\ m_{E_i}=\lambda_4c_iV+\Delta
m_i.$$

Please note that one also can add another symmetry (such as $U(1)$
or $Z_2$ ) into the model to let the gauge invariant term
$\overline{\varphi}_{iL} \rho_j E_{kR} $ absent from the Lagrangian
and $c_i=1$. Then the mass matrix and transform matrix will be more
simple.

Now let us consider the neutrino mass matrix. The right-handed
neutrino bare Majorana mass term is trivial, which is $M_N$ times
the identity. The Yukawa term $(\overline{\varphi}_{iL}N_{iR})\eta$
also contributes trivially to the Dirac mass matrix a term
proportional to the $3 \times 3$ identity matrix, {\it i.e.} $h_1u$
times the identity. The only non-trivial structure is from
contribution to the Dirac mass matrix supplied by the Yukawa
coupling to $\eta_i$, which is,
 \begin{eqnarray}\left(
\begin{array}{ccc}
0           & h_2<\eta_3>  &h_2<\eta_2>\\
h_2<\eta_3>  & 0            & h_2<\eta_1>\\
h_2<\eta_2>  & h_2<\eta_1>   & 0\\
\end{array}\right).
\end{eqnarray}

After making the assumption about $A_4$ breaking as,
\begin{equation}
<\eta_1>=u',\,\,\,\,\,\,<\eta_2>=<\eta_3>=0
\end{equation}

We obtain the full $6 \times 6$ neutrino mass matrix on the weak
interaction eigenstates,
\begin{eqnarray}
m_{\nu N}= \left(
\begin{array}{cccccc}
0     &0      &0     &h_1u &0      &0\\
0     &0      &0     &0       &h_1u  &h_2u'\\
0     &0      &0     &0       &h_2u'   &h_1u\\
h_1u  &0     &0      &M_N     &0       &0\\
0     &h_1u  &h_2u'   &0       &M_N     &0\\
0     &h_2u'  &h_1u  &0       &0       &M_N\\
\end{array}\right)
=\left(
\begin{array}{cc}
0   & m_D\\
m_D^T & m_S\\
\end{array}\right)
\end{eqnarray}
So the see-saw mass matrix for $(\nu_i)$ is,
\begin{eqnarray}
m_{\nu}=-m_Dm_s^{-1}m_D^T=-\frac{1}{M_N} \left(
\begin{array}{ccc}
(h_1u)^2&0                  &0\\
0        &(h_1u)^2+(h_2u')^2&2h_1h_2uu'\\
0        &2h_1h_2uu'        &(h_1u)^2+(h_2u')^2
\end{array}\right),\label{eq:mn}
\end{eqnarray}
which can be written as a simple form,
\begin{eqnarray}
m_{\nu}=-\frac{1}{M_N} \left(
\begin{array}{ccc}
\gamma   &0                  &0\\
0        &\alpha             &\beta\\
0        &\beta              &\alpha
\end{array}\right).\label{eq:mn1}
\end{eqnarray}
This mass matrix can be diagonalized by the transformation,
\begin{eqnarray}
V_{L}^\nu=\frac{1}{\sqrt{2}} \left(
\begin{array}{ccc}
0        &\sqrt{2}  &0\\
 1       &0         &-1\\
1        &0         &1
\end{array}\right).\label{eq:mn2}
\end{eqnarray}
And the mass matrix is given by,
\begin{eqnarray}
-\frac{1}{M_N} \left(
\begin{array}{ccc}
(h_1u+h_2u')^2&0                  &0\\
0        &(h_1u)^2                &0\\
0        &0                       &(h_1u-h_2u')^2
\end{array}\right).
\end{eqnarray}

Then we can choose the free parameters $h_i$s, $M_N$ and $u$ to get
the small masses of neutrinos. If $h_i$ is of order 1 and $M_N$ is
of order $\sim TeV$, $u, u'\sim MeV$, The neutrinos will have the
masses of order eV.

Useing the definition for the observed neutrino mixing matrix
$V=A_L^\dag V_L^\nu$ \cite{A4b,malatest}, we obtain the mixing
matrix in this order,
\begin{eqnarray}
V=A_L^\dag V_L^\nu = \left(
         \begin{array}{ccc}
         c_1& &\\
          &c_2 &\\
         & & c_3
\end{array}\right)\cdot
\left(
\begin{array}{ccc}
\frac{2}{\sqrt{6}}       &\frac{1}{\sqrt{3}}  &0\\
-\frac{1}{\sqrt{6}}      &\frac{1}{\sqrt{3}}  &\frac{i}{\sqrt{2}}\\
-\frac{1}{\sqrt{6}}      &\frac{1}{\sqrt{3}}  &-\frac{i}{\sqrt{2}}
\end{array}\right)=P_cV_{tri-bi}P_\phi \label{eq:v1}
\end{eqnarray}
where $V_{tri-bi}$ is defined in the Eq. (\ref{eq:1}), the phase
matrices $P_c$ and $P_\phi$ are both diagonal and with the diagonal
elements $c_i$ and the diagonal elements $1$, $1$ and $i$
respectively.

\section{Modified neutrinos masses and mixing matrix }\label{mod}

More generally, according to the low energy effective field theory
analysis \cite{o5} neutrino masses are generated by the unique
effective dimension-five operator $O_5$ which has been studied by
many papers in the standard $SU(2)_L\otimes U(1)_Y$ model
\cite{o5,neu} and $A_4$ model \cite{A4b, malatest} and has the form
as,
\begin{eqnarray}
O_5=\frac{\lambda_{ij}(H\varphi_L)_i^T(H'
\varphi_L)_j}{\Lambda}+h.c.
\end{eqnarray}

Note here the charge conjugation matrix $C$ between the lepton
fields has been omitted, and in our notation, $H$ and $H'$ denote
$\eta$ or $\eta_i$, $\Lambda$ denote $M_N$, $\varphi_L$ is the
$SU(3)$ lepton triplet and $\lambda$ is a matrix in flavour space.
Then there has three types of this $O_{5}$ operators, {\it i.e.},
$(\eta \varphi_{iL} )^{2},$ $(\eta \varphi_{iL} )(\eta_i
\varphi_{jL} )$ and $(\eta_i \varphi_{jL} )^{2}$. The operator
$(\eta \varphi_{iL} )^{2}$ which has the form of
$\underline{3}\times \underline{3}$ contributes a term proportional
to the identity matrix. Next, $(\eta \varphi_{iL} )(\eta_i
\varphi_{jL} ),$ which is formed by $ \underline{3}\times
\underline{3}\times \underline{3},$ has the form $\eta (\varphi
_{1L}\eta _{2}\varphi _{3L}+\varphi _{2L}\eta _{3}\varphi
_{1L}+\varphi _{3L}\eta _{1}\varphi _{2L}),$ and its analogous form.
Thus, the operator $ (\eta \varphi_{iL} )(\eta_i \varphi_{jL} )$
contributes the term denoted by $\beta $ in (\ref {eq:mn1}).
Finally, the operator $(\eta_i \varphi_{jL} )^{2}$ actually denotes
schematically 4 different operators since it is formed by
$(\underline{3} \times \underline{3})\times (\underline{3}\times
\underline{3})$ and this contains $\underline{1}\times
\underline{1}$, $\underline{1}^{\prime }\times \underline{1}^{\prime
\prime }$, $\underline{3}\times \underline{3}$, $\underline{3}\times
\underline{3}$ and $\underline{3}\times \underline{3}$,
corresponding respectively to the operators $(\eta _{1}\varphi
_{1L}+\eta _{2}\varphi _{2L}+\eta _{3}\varphi _{3L})^{2}$, $(\eta
_{1}\varphi _{1L}+\omega \eta _{2}\varphi _{2L}+\omega ^{2}\eta
_{3}\varphi _{3L})\cdot(\eta _{1}\varphi _{1L}+\omega ^{2}\eta
_{2}\varphi _{2L}+\omega \eta _{3}\varphi _{3L})$, $(\eta
_{2}\varphi _{3L},\eta _{3}\varphi _{1L},\eta _{1}\varphi
_{2L})\cdot(\eta _{3}\varphi _{2L},\eta _{1}\varphi _{3L},\eta
_{2}\varphi _{1L})$, $(\eta _{3}\varphi _{2L},\eta _{1}\varphi
_{3L},\eta _{2}\varphi _{1L})\cdot(\eta _{3}\varphi _{2L},\eta
_{1}\varphi _{3L},\eta _{2}\varphi _{1L})$ and $(\eta _{2}\varphi
_{3L},\eta _{3}\varphi _{1L},\eta _{1}\varphi _{2L})\cdot(\eta
_{2}\varphi _{3L},\eta _{3}\varphi _{1L},\eta _{1}\varphi _{2L})$.
Where $\eta$ and $\eta_i$ acquire the VEVs of $<\eta>=u$ and
$<\eta_i>=(u',0,0)$.

Then we obtain a more general form of the neutrino mass matrix
$m_\nu$ as,
\begin{equation}
m_{\nu }=-\frac{1}{M_N}\left(
\begin{array}{lll}
\gamma    & 0                     &  0  \\
0         &  \alpha -\varepsilon  & \beta\\
 0         & \beta                     & \alpha +\varepsilon
\end{array}
\right), \label{new}
\end{equation}
rather than the mass matrix $m_{\nu }$ in Eq. (\ref{eq:mn1})
\cite{A4c}.

At this point, it could only suppose that $\varepsilon $ is small
compared to $\beta ,$ in which case $U_{\nu }$ is perturbed from the
desired $V_L^{\nu } $ in Eq. (\ref{eq:mn2}) to

\begin{eqnarray}
V_{L}^\nu= \left(
\begin{array}{ccc}
0                 &1         &0\\
 cos\theta      &0         &-sin\theta\\
sin\theta        &0         &cos\theta
\end{array}\right).
\end{eqnarray}
where $\theta = \frac{\pi}{4} +\delta $ ( $\delta \ll 1$),
$sin\theta \simeq
\frac{1}{\sqrt{2}}(1+\frac{\varepsilon}{2\beta})$, $cos\theta
\simeq \frac{1}{\sqrt{2}}(1-\frac{\varepsilon}{2\beta})$. In this
order, the neutrino masses come out to be $\alpha +
\sqrt{\beta^{2}+\varepsilon^{2}} $, $\gamma$, and $\alpha
-\sqrt{\beta^{2}+\varepsilon ^{2}} $.

 Then the matrix V becomes,
\begin{eqnarray}
V=A_L^\dag V_L^\nu && \hspace{-0.5cm}= \frac{1}{\sqrt{3}}\left(
         \begin{array}{ccc}
         c_1& &\\
          &c_2 &\\
         & & c_3
\end{array}\right)\cdot
\left(
\begin{array}{ccc}
cos\theta+sin\theta                   &1  &cos\theta -sin\theta \\
\omega(sin\theta+\omega cos\theta )   &1  &\omega(cos\theta-\omega sin\theta )\\
\omega(cos\theta+\omega sin\theta )   &1 &\omega(-sin\theta+\omega
cos\theta )
\end{array}\right)\\
&&\hspace{-0.5cm}\simeq  \frac{1}{\sqrt{6}}\left(
         \begin{array}{ccc}
         c_1& &\\
          &c_2 &\\
         & & c_3
\end{array}\right)\cdot\left(
\begin{array}{ccc}
2                                         &\sqrt{2}  &-\frac{\varepsilon}{\beta} \\
-1+\frac{\sqrt{3}\varepsilon }{2\beta} i  &\sqrt{2}  &\sqrt{3}i+\frac{\varepsilon}{2\beta}\\
-1-\frac{\sqrt{3}\varepsilon }{2\beta}i   &\sqrt{2}
&-\sqrt{3}i+\frac{\varepsilon}{2\beta}
\end{array}\right)\\
&=&P_cV_{tri-bi}\left(\begin{array}{ccc}
         cos\delta & 0& -sin\delta\\
         0 &1 &0\\
       i sin\delta  & 0& i cos\delta
\end{array}\right)\label{v2}
\end{eqnarray}

Compared with Eq. (\ref{eq:v1}), the middle column is uncorrected at
this level. At the same time a nonzero $V_{e3}$ element is
generated, and there are other small deviations from exact
tri-bimaximal mixing.

\section{conclusion}\label{con}
The non-abelian discrete symmetry $A_4$ appears to be particularly
fit for the purpose to produce the neutrino tri-bimaximal mixing
pattern in the standard $SU(2)_L \times U(1)_Y$ theory. In this
paper we have generalized the $A_4$ study to the 3-3-1 model. In the
version we consider here there are negative charged leptons and
right handed neutrinos in addition to the ordinary SM leptons. By
combining the $A_4$ symmetry and canonical see saw mechanism we have
reproduced the observed neutrino tri-bimaximal mixing matrix. The
smallness of neutrino masses can be explained naturally without
introducing too heavy neutral leptons. Our main results are Eq.
(\ref{eq:v1}) and Eq. (\ref{v2}). Numerically they are consistent
with the present experimental constraints \cite{data,fit,GG03}. A
small but non-zero mixing angle $V_{e3}$ present in Eq. (29). This
angle has been assumed to be zero in tri-bimaximal form but it is
only required experimentally to be small {\it i.e.} $|V_{e3}|<
0.16$\cite{mf}, and maybe is measured more accurately by the daya
bay reactor neutrino experiments \cite{dayabay}.

\section*{Acknowledgments}
We would like to thank Professor Xiaoyuan Li for suggesting this
study, carefully reading of the manuscript and valuable guidance.
This work is supported in part by the China National Natural Science
Foundation under grant No. 10475106.


\section*{Appendix: Basic $A_4$ properties and the potential }

The model is based on the discrete group $A_4$ following refs.
\cite{max3,malast},
 where its structure and representations are described in detail. It is appropriate to
recall briefly some relevant features of it. $A_4$ symmetry is the
discrete symmetry group of the rotations that leave a
 tethraedron invariant, or the group of the even permutations of 4 objects.
 It has 12 elements and 4 inequivalent irreducible representations denoted $\underline{1}$, $\underline{1}'$, $\underline{1}''$
 and $\underline{3}$  in terms of their respective dimensions. Introducing $\omega$, the cubic root of
 unity, $\omega=\exp{i\frac{2\pi}{3}}$, so that $1+\omega+\omega^2=0$, the three
 one-dimensional representations are obtained by dividing the 12 elements of $A_4$
 in three classes, which are determined by the multiplication rule, and assigning
 to (class 1, class 2, class 3) a factor $(1,1,1)$ for $\underline{1}$, or $(1,\omega,\omega^2)$
 for $\underline{1}'$ or $(1,\omega^2,\omega)$ for $\underline{1}''$. The product of two $\underline{3}$ gives
 $\underline{3} \times \underline{3} = \underline{1} + \underline{1}' + \underline{1}'' + \underline{3} + \underline{3}$. Also $\underline{1}' \times \underline{1}' = \underline{1}''$,
 $\underline{1}' \times \underline{1}'' = \underline{1}$, $\underline{1}'' \times \underline{1}'' = \underline{1}'$ etc.
For $\underline{3}\sim (a_1,a_2,a_3)$ and $\underline{3}\sim
(b_1,b_2,b_3)$, the irreducible representations obtained from their
tensor products,
\begin{eqnarray}\nonumber
\underline{3} \times \underline{3} &=&
\underline{1}(a_1b_1+a_2b_2+a_3b_3) + \underline{1}'(a_1b_1+\omega
a_2b_2+\omega^2 a_3b_3)+ \underline{1}''(a_1b_1+\omega^2
a_2b_2+\omega a_3b_3) \\
&&+ \underline{3}(a_2b_3,a_3b_1,a_1b_2) +
\underline{3}(a_3b_2,a_1b_3,a_2b_1)
\end{eqnarray}

As required in section \ref{a4}, there are need a mechanism such
that the scalar fields develop a VEV along the directions,
\begin{eqnarray}\nonumber
&&<\chi> \,=\, V,\,\,\,\,\\\nonumber
\nonumber
&&<\rho_i> \,=\,
(v,v,v),\,\,\,\,\\\nonumber
&& <\eta>\, =\,u,\,\,\,\,\\
&&<\eta'_i>\,=\, (u',0,0).\label{eq:vev}
\end{eqnarray}
Here for the convenience of depiction, we replace $\eta_i$ with
$\eta'_i$ in this section.

When we study the Higgs potential for $SU(3)\times U(1)$ scalars, we
should consider all various representations for $A_{4}.$ For the
sake of simplicity, previous to give the whole potential, we
restrict to the Higgs potential
 for a single $SU(3)\times U(1)$ scalar triplet $\phi_i $
which transform $\underline{3}$ under $A_{4}.$ The multiplication
$ \underline{3}\times
\underline{3}=\underline{1}+\underline{1}^{\prime }+
\underline{1}^{\prime \prime }+\underline{3}+\underline{3}$ shows
 that there is only one quadratic invariant.  Since
($\underline{3}\times \underline{3})\times (\underline{3}\times
\underline{3})$ contains $\underline{1}$ five times, corresponding
to $ \underline{1}\times \underline{1},$ $\underline{1}^{\prime
}\times \underline{1}^{\prime \prime },$ $\underline{3}\times
\underline{3},$ $\underline{3}\times \underline{3},$ and $
\underline{3}\times \underline{3},$ there should have 5 quartic
invariants and the last two terms $\underline{3}\times
\underline{3}$ and $\underline{3}\times \underline{3}$ are the
complex conjugate with each other. Then the Higgs potential of the
single $SU(3)\times U(1)$ scalar triplet $\phi_i $ which transform
$\underline{3}$ under $A_{4}$ is given by,
\begin{eqnarray}
V(\phi) &=& \mu_\phi^2 \sum_i \phi_i^\dagger \phi_i + {1 \over 2}
\lambda_1 ( \sum_i \phi_i^\dagger \phi_i )^2 + {1 \over 2}
\lambda_2 \sum_{i,j} (3 \delta_{i,j} - 1) (\phi_i^\dagger
\phi_i)(\phi_j^\dagger \phi_j) \nonumber
\\ && +~ {1 \over 2} \lambda_3 \sum_{i \neq j} (\phi_i^\dagger \phi_j)
(\phi_j^\dagger \phi_i) + {1 \over 2} \lambda_4 \sum_{i \neq j}
(\phi_i^\dagger \phi_j)^2.
\end{eqnarray}
Note that for the sake of simplicity, we have taken $\lambda$'s
and $v_i$'s to be real, since our focus here is not on $CP$
violation. This potential has minimum at
$v_1=v_2=v_3=v=\sqrt{-\frac{\mu_\phi^2}{3\lambda_1+2\lambda_3+2\lambda_4}}$
or at
$v_1=\sqrt{-\frac{\mu_\phi^2}{\lambda_1+2\lambda_2}},\,\,v_2=v_3=0$.

Consider all the scalar triplets in the model and note that under
$A_4$ the $\underline{3}\times
\underline{3}\times\underline{3}=\underline{1}$ is possible, {\it
i.e.} $1$ $2$ $3$ $+$ permutations. So the $SU(3)_C \otimes
SU(3)_L\otimes U(1)_X\otimes A_4$ invariant higgs potential of the
model can be written as,
\begin{eqnarray}
 V(\chi)&=&\mu_\chi^2(\chi^\dag
           \chi)+\frac{1}{2}\lambda_1^\chi(\chi^\dag \chi)^2\\
V(\rho)& =& \mu_\rho^2 \sum_i \rho_i^\dagger \rho_i + {1 \over 2}
       \lambda_1^\rho ( \sum_i \rho_i^\dagger \rho_i )^2 + {1 \over 2}
           \lambda_2^\rho \sum_{i,j} (3 \delta_{i,j} - 1) (\rho_i^\dagger
           \rho_i)(\rho_j^\dagger \rho_j) \nonumber
\\ && +~ {1 \over 2} \lambda_3^\rho \sum_{i \neq j} (\rho_i^\dagger \rho_j)
          (\rho_j^\dagger \rho_i) + {1 \over 2} \lambda_4^\rho \sum_{i \neq j}
             (\rho_i^\dagger \rho_j)^2.
\\ V(\eta)&=&\mu_{\eta}^2(\eta^\dag
           \eta)+\frac{1}{2}\lambda_1^{\eta}(\eta^\dag \eta)^2\\
V(\eta')& =& \mu_{\eta'}^2 \sum_i \eta_{i}'^{\dagger} \eta'_i + {1
\over
        2}\lambda_1^{\eta'} ( \sum_i \eta_i'^{\dagger} \eta'_i )^2 + {1 \over 2}
           \lambda_2^{\eta'} \sum_{i,j} (3 \delta_{i,j} - 1)
            (\eta_i'^{\dagger}
           \eta'_i)(\eta_j'^{\dagger} \eta'_j) \nonumber
\\ && + {1 \over 2} \lambda_3^{\eta'} \sum_{i \neq j} (\eta_i'^{\dagger} \eta_j)
         (\eta_j'^{\dagger} \eta'_i) + {1 \over 2} \lambda_4^{\eta'} \sum_{i \neq j}
             (\eta_i'^{\dagger} \eta'_j)^2.\\\nonumber
V(\chi\rho)&=&\lambda_1^{\chi\rho}(\chi^\dag
              \chi)(\sum_i\rho_i^\dag \rho_i)+\lambda_2^{\chi\rho}\sum_i(\chi^\dag
             \rho_i )(\rho_i^\dag \chi)+{1 \over 2}\lambda_3^{\chi\rho}\sum_i[(\chi^\dag
             \rho_i )(\chi^\dag \rho_i)+h.c]\\
            && +\lambda_4^{\chi\rho}|\epsilon_{ijk}|[(\chi^\dag\rho_i )(\rho_j^\dag \rho_k)+h.c.]\\
V(\chi\eta)&=&\lambda_1^{\chi\eta}(\chi^\dag
              \chi)(\eta^\dag \eta)+\lambda_2^{\chi\eta}(\chi^\dag
             \eta )(\eta^\dag \chi)+\mu^{\chi\eta}(\chi\chi\eta)\\
V(\chi\eta')&=&\lambda_1^{\chi\eta'}(\chi^\dag
              \chi)(\sum_i\eta_i'^\dag \eta'_i)+\lambda_2^{\chi\eta'}\sum_i(\chi^\dag
             \eta'_i )(\eta_i'^\dag \chi)\\
V(\rho\eta)&=&\lambda_1^{\rho\eta} \sum_i(\rho_i^\dag
              \rho_i)(\eta^\dag \eta)+\lambda_2^{\rho\eta}\sum_i(\rho_i^\dag
             \eta )(\eta^\dag \rho_i)+\mu^{\rho\eta}\sum_i(\rho_i\rho_i\eta)\\\nonumber
V(\rho\eta')&=&\lambda_1^{\rho\eta'}(\sum_i\rho_i^\dag
              \rho_i)(\sum_i\eta_i'^\dag \eta'_i)
              +{1\over 2}\lambda_2^{\rho\eta'}\sum_{i,j}(3 \delta_{i,j} - 1)(\rho_i^\dag
             \rho_i )(\eta_j'^\dag \eta'_j)\\\nonumber
       && + {1 \over 2} \lambda_3^{\rho\eta'} \sum_{i \neq j} (\rho_i^\dagger \rho_j)
         (\eta_j'^\dagger \eta'_j) + {1 \over 2} \lambda_4^{\rho \eta'} \sum_{i \neq j}(\rho_i^\dagger \rho_j)
         (\eta_i'^\dagger \eta'_j)+{1\over 2}\lambda_5^{\rho\eta'}\sum_{i,j}(3 \delta_{i,j} - 1)(\rho_i^\dag
             \eta_i )(\eta_j'^\dag \rho'_j)\\
       && + {1 \over 2} \lambda_6^{\rho\eta'} \sum_{i \neq j} (\rho_i^\dagger \eta'_j)
         (\eta_j'^\dagger \rho_i) + {1 \over 2} \lambda_7^{\rho\eta'} \sum_{i \neq j}(\rho_i^\dagger \eta'_j)
         (\eta_i'^\dagger
         \rho_j)+\mu^{\rho\eta'}|\varepsilon^{ijk}|(\rho_i\rho_j\eta'_k)\\\nonumber
V(\eta\eta')&=&\lambda_1^{\eta\eta'}(\eta^\dag
              \eta)(\sum_i\eta_i'^\dag \eta'_i)+\lambda_2^{\eta\eta'_i}\sum_i(\eta^\dag
             \eta'_i )(\eta_i'^\dag \eta)+{1 \over 2}\lambda_3^{\eta\eta'}\sum_i[(\eta^\dag
             \eta'_i )(\eta^\dag \eta'_i )+h.c]\\
      && +\lambda_4^{\eta\eta'}|\epsilon_{ijk}|[(\eta^\dag\eta'_i )(\eta_j'^\dag \eta'_k)+h.c.]\\
V(\chi\rho\eta')&=&\lambda^{\chi\rho\eta'}_1|\epsilon_{ijk}|[(\chi^\dag\rho_i
               )(\eta_j'^\dag\eta'_k)+h.c.]
               +\lambda^{\chi\rho\eta'}_2|\epsilon_{ijk}|[(\chi^\dag\eta'_i
               )(\eta_j'^\dag \rho'_k)+h.c.]+
          \mu^{\chi\rho\eta'}\sum_i\rho_i\eta'_j\chi\\\nonumber
V(\chi\rho\eta\eta')&=&\lambda_1^{\chi\rho\eta\eta'}\sum_i(\chi^\dag\rho_i)(\eta^\dag\eta'_i)+
                    \lambda_2^{\chi\rho\eta\eta'}\sum_i(\chi^\dag\rho_i)(\eta_i'^\dag\eta)
                  +\lambda_3^{\chi\rho\eta\eta'}\sum_i(\chi^\dag\eta)(\eta_i'^\dag\rho_i)\\
                  &&+\lambda_4^{\chi\rho\eta\eta'}\sum_i(\chi^\dag\eta'_i)(\eta^\dag\rho_i)+h.c
\end{eqnarray}

Note that, in the order we discussed, the charged lepton mass matrix
and the neutrino mass matrix are related by two separate scalar
sectors $\chi$, $\rho_i$ and $\eta$, $\eta'_i$, respectively. If
there is no communication between the two scalar sectors, the
 VEVs of $\rho$ and $\eta'_i$ given in Eq. (\ref{eq:vev})
 will break the $A_4$ symmetry to $Z_3$ symmetry in
charged sector and $Z_2$ symmetry in neutrino sector and these
residual symmetries will be maintained. In general, $\chi$, $\rho_i$
and $\eta$, $\eta'_i$ mix in the potential and it is not possible to
keep the VEVs structure for $\rho$ and $\eta'_i$ as Eq.
(\ref{eq:vev}). One needs to separate them from communicating in the
scalar potential and therefore to simplify the vacuum alignment
problem. Suppose that, at least at some level, the interchange
between the fields $\chi, \rho_i$ and $\eta, \eta'_i$ to produce the
desired mass matrices in the charged and neutrino lepton sectors. We
can determine the minima of two scalar potentials $V_c$ and $V_0$,
depending only, respectively, on $\chi$, $\rho_i$ and $\eta$,
$\eta'_i$. There are whole regions of the parameter space where
$V_c(\chi, \rho)$ and $V_0(\eta, \eta')$ have the minima VEVs given
in Eq. (\ref{eq:vev}).

First consider the scalar potential $V_c(\chi,\rho)$,
\begin{eqnarray}
V_c(\chi,\rho)=V(\chi)+V(\rho)+V(\chi\rho).
\end{eqnarray}
Analyzing the field configuration as,
\begin{eqnarray}
<\chi>=V,\,\,\,\,\ <\rho>=(v,v,v),
\end{eqnarray}
then the minimum conditions are,
\begin{eqnarray}
\frac{\partial V_c}{\partial \chi}&=&2V(\mu_\chi^2+\lambda_1^\chi
V^2+3\lambda_1^{\chi\rho}v^2)\\
 \frac{\partial V_c}{\partial\rho_i}&=&2v(\mu_\rho^2+3\lambda_1^\rho
                      v^2+2\lambda_3^\rho v^2+2\lambda_4^\rho
                     v^2+\lambda_1^{\chi\rho}V^2)
\end{eqnarray}

Therefore the $<\chi>=V$  and $<\rho_i>=(v,v,v)$ can be local minima
of $V_c$ depending on the parameters. Take into account $A_4$
symmetry there are four degenerate minima, {\it i.e.},
$<\rho_i>=(v,v,v)$, $<\rho_i>=(v,-v,-v)$, $<\rho_i>=(-v,v,-v)$ and
$<\rho_i>=(-v,-v,v)$ in this region.

Then consider the scalar potential $V_0(\eta,\eta')$,
\begin{eqnarray}
V_0(\eta,\eta')=V(\eta)+V(\eta')+V(\eta\eta').
\end{eqnarray}
Search for the minimum conditions at $<\eta>=u$,
$<\eta'_i>=(u',0,0)$, then
\begin{eqnarray}
\frac{\partial V_0}{\partial
\eta}&=&2u(\mu_{\eta}^2+\lambda_1^{\eta}
u^2+\lambda_1^{\eta\eta'}u'^2+\lambda_2^{\eta\eta'}u'^2+\lambda_3^{\eta\eta'}u'^2)\\
\frac{\partial
 V_0}{\partial\eta'_1}&=&2u'(\mu_{\eta'}^2+\lambda_1^{\eta'}
                      u'^2+2\lambda_2^{\eta'_i} u'^2+\lambda_1^{\eta\eta'}u^2
                      +\lambda_2^{\eta\eta'}u^2+\lambda_3^{\eta\eta'}u^2)
\end{eqnarray}
In this case $(\partial V_0/\partial \eta'_{2,3})=0$ are
automatically satisfied. There a large portion of the parameter
space where the minimum is. In this region, there are six degenerate
minima, {\it i.e.}, $<\eta'_i>=(\pm u',0,0)$, $<\eta'_i>=(0,\pm
u',0)$ and $<\eta'_i>=(0,0,\pm u')$, related by $A_4$ symmetry.
Putting together the minima of $V_c(\chi,\rho)$ and
$V_0(\eta,\eta')$ there are 24 degenerate minima of the potential
energy, differing for signs or ordering. It can be shown that these
24 minima produce exactly the same mass pattern discussed in section
\ref{a4}, up to field and parameter redefinitions. Therefore it is
not restrictive to choose one of them.


\begin{thebibliography}{99}



\bibitem{nos}For a brief review, see A. Yu. Smirnov, hep-ph/0402264;
            G. Altarelli, Nucl. Phys. {\bf B}, Proc. Suppl. 143,
            470(2005).

\bibitem{superK} Super-Kamiokande Collaboration, Y. Fukuda {\it et
                   al}., Phys. Rev. Lett. {\bf 81}, 1562(1998); Y. Ashie {\it et
                    al}, Phys. Rev. Lett. {\bf 93}, 101801(2004);
               hep-ex/0510064.

\bibitem{kam} KamLAND Collaboration, K. Eguchi {\it et al.},  Phys.
              Rev. Lett. {\bf 90}, 021802(2003); T. Araki {\it et al}, Phys.
              Rev. Lett. {\bf 94}, 081801(2005).

\bibitem{sno} SNO Collaboration, Q. R. Ahmad {\it et al.}, Phys.
              Rev. Lett. {\bf 89}, 011301(2002);  Phys. Rev. Lett. {\bf 89},
               011302 (2002);  Phys. Rev. Lett. {\bf 92}, 181301(2004); B.
               Aharmim {\it et al}, Phys. Rev.  {\bf C72}, 055502(2005).


\bibitem{V1}P. F. Harrision, D. H. Perkins and W. G. Scott, Phys. Lett. {\bf B530}, 167
(2002); Z. Z. Xing, Phys. Lett. {\bf B533}, 85(2002).

\bibitem{V2}X. G. He and A. Zee, Phys. Lett. {\bf B560}, 87(2003); Phys. Rev. {\bf
             D68}, 037302(2003).


\bibitem{o5}  S. Weinberg, Phys. Rev. Lett. {\bf 43}, 1566(1979);
F. Wilczek and A. Zee, Phys. Rev. Lett. {\bf 43}, 1571(1979).


\bibitem{neu} E. Ma Phys. Lett. {\bf 81}, 1171(1998);
 Phys. Rev. Lett. {\bf 86}, 2502 (2001);
     Phys. Rev. {\bf D66}, 037301(2002); Phys. Rev. {\bf D73},
     077301(2006).


\bibitem{seesaw} M. Gell-Mann, P. Ramond, and R. Slansky, in
{\em Supergravity}, edited by P. van Nieuwenhuizen and D. Z.
Freedman (North-Holland, Amsterdam, 1979), p.~315; T. Yanagida, in
{\em Proceedings of the Workshop on the Unified Theory and the
Baryon Number in the Universe}, edited by O. Sawada and A.
Sugamoto (KEK Report No.~79-18, Tsukuba, Japan, 1979), p.~95; R.
N. Mohapatra and G. Senjanovic, Phys. Rev. Lett. {\bf 44}, 912
(1980).

\bibitem{tribim} C.
I. Low and R. R. Volkas, Phys. Rev. {\bf D68}, 033007(2003);  W.
Rodejohann and Z. Z. Xing, Phys. Lett. {\bf B601}, 176(2004); Z.
Z. Xing and S. Zhou, Phys. Lett. {\bf B606}, 145(2005); J. W. Mei,
Z. Z. Xing, Phys. Lett. {\bf B623}, 227(2005); J. E. Kim and J.-C.
Park, {\bf JHEP0605}, 017(2006); W. Grimus and L. Lavoura, {\bf
JHEP0601}, 018(2006); I. Varizelas, S.-F. King and G. G. Ross,
hep-ph/0512313, hep-ph/0607045; N. Singh, M. Rajkhowa and A.
Borach, hep-ph/0603189; P. Kovtun and A. Zee, Phys. Lett. {\bf
B640}, 37(2006); R. Mohapatra, S. Naris and Y.-H. Yu, Phys. Lett.
{\bf B639}, 318(2006); N. Haba, A. Watanabe and K. Yoshioka, Phys.
Rev. Lett. {\bf 97}, 041601(2006).

\bibitem{A4a}
          E. Ma, Mod. Phys. Lett. {\bf A17}, 289 (2002); ibid. {\bf A17},
           627(2002); E. Ma, Mod. Phys. Lett. {\bf A17}, 2361(2002).

\bibitem{A4b}  E.
          Ma, hep-ph/0208077; hep-ph/0208097; hep-ph/0307016;
           hep-ph/0311215; New
           J. Phys. {\bf 6}, 104(2004); hep-ph/0409075; E. Ma, Mod. Phys.
         Lett. {\bf A20}, 1953(2005); S. L. Chen, M. Frigerio, and E.
         Ma, Nucl. Phys. {\bf A764}, 423(2006); E. Ma,
         hep-ph/0606024; hep-ph/0606039.

\bibitem{A4c} E. Ma, Phys. Rev. {\bf D66}, 117301(2002); Phys. Rev.  {\bf
D73}, 057304(2006); K.~S.~Babu and X. G.~He, hep-ph/0507217;
         G. Altarelli and F. Feruglio, Nucl. Phys. {\bf B720},
64(2005); Nucl. Phys. {\bf B741}, 215(2006); K. S. Babu and X-G.
He, hep-ph/0507217.



\bibitem{A4hv} X. G. He, Yong-Yeon Keum and Ray Volkas,
{\bf JHEP0604}, 039(2006);  X.-G. He and A. Zee, hep-ph/067163; S.
K. Kang, Z. Z. Xing and S. Zhou, Phys. Rev. {\bf D73},
013001(2006).



\bibitem{pisano} F. Pisano and V. Pleitez, Phys. Rev. {\bf D46}, 410(1992);
         P. H. Frampton, Phys. Rev. Lett. {\bf 69},
                 2889(1992).


\bibitem{foot} R. Foot, H. N. Long and T. A. Tran, Phys. Rev. {\bf D50},
       R34(1994); H. N. Long, Phys. Rev. {\bf D53}, 437(1996);
        \textit{ibid}, \textbf{D54}, 4691(1996); H. N. Long, Mod. Phys.
       Lett. \textbf{A13}, 1865(1998). A. Carcamo, R. Martinez and
    F.Ochoa, Phys. Rev. \textbf{D73}, 035007(2006); D. Chang and H. N.
    Long, Phys. Rev. {\bf D73}, 053006(2006).

\bibitem{bb} R. A. Diaz R. Martinea and F. Ocha, Phys. Rev.  {\bf D69},
095009(2004); Phys. Rev.  {\bf D72}, 035018(2005).



\bibitem{long}  H. N. Long and T. A. Tran, Mod. Phys.
             Lett. {\bf A9}, 2507(1994);  F. Pisano and V. Pleitez, Phys.
          Rev. {\bf D51}, 3865(1995);  I. Cotaescu, Int. J. Mod. Phys.
           {\bf A12}, 1483(1997); A. Doff and F. Pisano, Phys. Rev.
           {\bf D63}, 097903(2001); hep-ph/0011087.

\bibitem{nc}J. Alexis Rodriguez and M. sher, Phys. Rev.  {\bf D70},
117702(2004).




\bibitem{montero}  J. C. Montero, F.
              Pisano and V. Pleitez, Phys. Rev. {\bf D47}, 2918(1993); R. Foot,
             O. F. Hernandez, F. Pisano and V. Pleitez, Phys. Rev. {\bf D47},
          4158(1993); V. Pleitez and M. D. Tonasse, Phys. Rev. {\bf D48},
             2353(1993); {\it ibid} 5274(1993);
              L. Epele, H. Fanchiotti, C. Garc\'\i a Canal and D.
             G\'omez Dumm, Phys. Lett. {\bf B343}, 291(1995); M. \"Ozer, Phys.
          Rev. {\bf D54}, 4561(1996).

\bibitem{tonasse}M. D. Tonasse, Phys. Lett. {\bf B381}, 191(1996); D. Gomez Dumm,
         Int. J. Mod. Phys. {\bf A11}, 887(1996); N. T. Anh, N. A. Ky,
             and H. N. Long, Int. J. Mod. Phys. {\bf A15}, 283(2000);
             {\it ibid}, {\bf A16}, 541(2001).

\bibitem{singer}M. Singer, J. W. F. Valle and J. Schechter, Phys. Rev. {\bf D22},
          738(1980);
       M. \"Ozer, Phys. Rev. {\bf D54}, 1143(1996).

\bibitem{la}L. A. S\'anchez, W. A. Ponce, and R. Mart\'\i nez, Phys. Rev. {\bf
        D64}, 075013(2001);
       R. Mart\'\i nez, W. A. Ponce and L. A. S\'anchez, Phys. Rev. {\bf
       D65}, 055013(2002).
           W. A. Ponce, J. B. Fl\'orez,
         and L. A. S\'anchez, Int. J. Mod. Phys. {\bf A17}, 643(2002).





\bibitem{yasue}  Y. Okamoto and M.
Yasu${\grave {\rm e}}$, Phys. Lett. {\bf B466}, 267(1999); T.
Kitabayashi and M. Yasue, Phys. Rev. {\bf D63}, 095002(2001); Phys.
Rev. {\bf D63}, 095006(2001); Nucl. Phys. {\bf B609}, 61(2001);
Phys. Rev. {\bf D67}, 015006(2003).

\bibitem{Lately}
    M. B. Tully and G. C. Joshi, Phys. Rev. {\bf D64}, 011301(R)(2001);
    J. C. Montero, C. A. Pires and V. Pleitez, Phys. Lett. {\bf B502},
    167(2001).



\bibitem{wa}W. A. Ponce, Y. Giraldo and L. A. Sanchez, hep-ph/0201133;
           W. A. Ponce, Y. Giraldo and L. A. Sanchez, Phys. Rev.
  {\bf D67}, 075001(2003).



\bibitem{kita}T. Kitabayashi and M. Yasue, Phys. Lett. \textbf{B508},
85(2001); Phys. Rev. {\bf D63}, 095002(2001); Phys. Rev. {\bf
D63}, 095006(2001); T. Kitabayashi, Phys. Rev. {\bf D64},
057301(2001); hep-hp/0010341.


\bibitem{pires}J. C. Montero, C. A. de S. Pires and V. Pleitez, Phys. Rev.
\textbf{D65}, 093017(2002); J. C. Montero, C. A. de S. Pires and
V. Pleitez, Phys. Rev. \textbf{D65}, 095001(2002); J. C. Montero,
C. A. de S. Pires and V. Pleitez, Phys. Rev. \textbf{D66},
113003(2002).

\bibitem{dng} D. Ng, Phys. Lett {\bf D49},
4805(1994); A. Palcu, Mod.
             Phys. Lett {\bf A21}, 1203(2006).







\bibitem{malast}
E.~Ma,  Phys. Rev.  {\bf D70}, 031901(R)(2004); E.~Ma,
arXiv:hep-ph/0409075; E.~Ma,
  New J.\ Phys.\  {\bf 6}, 104(2004).
  G. Alatarelli and F. Feruglio, Phys. Rev.  {\bf D72}, 094030(2005).

\bibitem{max3}
E.~Ma and G.~Rajasekaran,
  Phys. Rev.  {\bf D64}, 113012(2001);
K.~S.~Babu, E.~Ma and J.~W.~F.~Valle, Phys. Lett. {\bf B552},
 207(2003); M.~Hirsch, J.~C.~Romao, S.~Skadhauge, J.~W.~F.~Valle
and A.~Villanova del Moral, hep-ph/0312244; M.~Hirsch, J.~C.~Romao,
S.~Skadhauge, J.~W.~F.~Valle and A.~Villanova del Moral, Phys. Rev.
{\bf D69}, 093006(2004).

\bibitem{gtom}  P. Langacker and D. london, Phys. Rev. {\bf
             D38}, 886(1988).


\bibitem{malatest} A.~Zee,
       Phys. Lett. {\bf B630}, 58(2005); G. Altarallia, hep-ph/0508053;  E.~Ma, Phys. Rev. {\bf
        D72}, 037301(2005); Moden. Phys. Lett {\bf A20}, 2601(2005);
       X. G. He, Y. Y. Keum and Raymond R. Volkas, JHEP{\bf0604}, 039(2006).

\bibitem{data} MARCO Collaboration, M. Ambrosio et al., Eur. Phys. J. {\bf C36},
          323(2004); Soudan 2 Collaboration, M. Sanchez et al., Phys. Rev.
          {\bf D68}, 113004(2003); CHOOZ Collaboration, M. Apollonio et al.,
          Eur. Phys. J. {\bf C27}, 331(2003); S. Eidelman, et al., Particle
          Data Group, Phys. Lett. {\bf B592}, 1(2004);  K2K Collaboration, E.
          Aliu et al., Phys. Rev. Lett. {\bf 94}, 081802(2005).

\bibitem{fit} A. Strumia and F. Vissani, hep-ph/0606054; X. G. He
and A. Zee, hep-ph/0607163.



\bibitem{GG03} C. Gonzalez-Garcia, {\tt http://www.dpf2003.org/xx/ neutrino/concha.pdf}.

\bibitem{mf} M. Apollonio, et al., Phys. Lett. {\bf B466},
415(1999); F. Boehm, et al., Phys. Rev. {\bf D64}, 112001(2001).


\bibitem{dayabay} For the brief introduction of the experiments, see
http://dayawane.ihep.ac.cn/cn/index.html.




\end{thebibliography}
\end{document}